# Floating zone growth of high-purity MgO substrate single crystals

**Christo Guguschev, [a,*] Michael Schulze,[a] Andrea Dittmar,[a] Detlef Klimm,[a] Kaspars Dadzis,[a] Thomas Schroeder,[a,b] Knut Peters,[c] Aakash Pushp [d,*]**

[a] Leibniz-Institut für Kristallzüchtung, Max-Born-Str. 2, 12489 Berlin, Germany

[b] Humboldt-Universität zu Berlin, Institut für Physik, Newtonstr. 15, 12489 Berlin, Germany

[c] CrysTec GmbH, Köpenicker Str. 325, 12555 Berlin, Germany

[d] IBM Almaden Research Center, 650 Harry Road, San Jose, CA 95120, USA

**Synopsis** High-purity MgO substrate crystals grown by the optical floating zone technique are demonstrated, which can serve as a literal foundation for the development of novel device concepts based on epitaxially grown thin films.

**Abstract** MgO single crystals with diameters between 3.5 and 5 mm and lengths up to 40 mm were grown by the optical floating zone technique (OFZ). Despite challenging material properties such as the high melting point of 2825 °C, very high evaporation rate and perfect {100} cleaving characteristics, crack-free crystals were grown at high growth rates exceeding 40 mm/h and at high thermal gradients. Chemical investigations revealed that the OFZ technique is suitable for the preparation of substrate crystals with a purity of 5N to facilitate the development of novel demonstrator devices based on epitaxially grown thin films. The achieved purity level is improved by more than one order of magnitude if compared to commercial MgO substrate single crystals graded as high purity.

[*] Corresponding authors. E-mail address: christo.guguschev@ikz-berlin.de (C. Guguschev).
apushp@us.ibm.com, apushp@gmail.com (A. Pushp).

1. Introduction

Melt-grown MgO single crystals are commercially available and most of these crystals are grown on a very large scale using the arc-fusion technique. In terms of the incorporation of impurities like Ca, Al and various transition metals in such crystals, this technique seems to be limited to crystals with a total purity of about 99.99%, since beyond that level contamination from the electrodes and other sources are dominant (Butler *et al.*, 1971, Freund *et al.*, 1978). In a more recent study, electrodes made from high-purity graphite were used, but impurity levels were not presented (Zhang *et al.*, 2005). Chemical vapor deposition is an alternative growth technique for the growth of MgO crystals, where crystals with a cation impurity of about 150 ppm were demonstrated (Booth *et al.*, 1975).

In this study, we demonstrate that the optical floating zone technique can be utilized to obtain high-purity MgO crystals significantly purer than commercial crystals. To the best of our knowledge, this is the first demonstration that MgO crystals can be successfully grown using this technique. Such high-purity crystals shaped into well-polished substrates can serve as a literal foundation for the development of novel device concepts based on epitaxially grown thin films.

2. Material and methods

    2.1 Starting materials and crystal growth

The ⟨100⟩-oriented cylindrical seeds used were machined from commercial arc-fusion grown crystals. Similarly prepared rods with the same orientation were used as initial feeds. Subsequently, concentrated $HNO_3$ was used as an etchant to remove potential residuals from the surface of the rods. Rods with diameters of about 5 mm and typical lengths in the range 50-65 mm were prepared. Additionally, feed rods with similar dimensions were machined from high-purity ceramics (4N5) supplied by Kojundo.

To increase the purity further, the route of the classic ceramic preparation technique was chosen. For this preparation procedure, 5N purity powder (Roth, Rotimetric 5N) was filled in rubber balloons and cold-isostatically pressed at 2000 bar. Subsequently, the green bodies were sintered in a conventional muffle furnace equipped with $MoSi_2$ heating elements in air. The rods were enclosed in a platinum boat with a lid. The temperatures and holding times were as follows: 2 h at 1200 °C followed by 5 h at 1550 °C. The applied heating ramps were 100 K/h and the cooling rate was 200 K/h.

The crystal growth experiments were performed using a conventional image furnace with a vertical double ellipsoid configuration (model HKZ10, SciDre). A 6.5 kW Xenon short-arc lamp was used as a radiation source. Translation directions of the feeding and pulling shafts were downwards and no rotation of the growing crystal was applied. Most of the crystals were grown at growth rates in the range of 20 to 70 mm/h. Relevant crystal growth parameters are summarized in Table 1 for various crystals. For the first growth experiments, fused silica tubes with outer diameters of about 36 and 20 mm were used as a growth chamber and protection tube, respectively. The gas flow within this initial standard setup (V.1) was directed downwards. For the updated growth setup (V.2), the gas flow direction was upwards, and the diameters of the fused silica tubes were increased to 69 mm and 55 mm, respectively.

**Table 1** Summary of the crystal growth conditions.

| Crystal | Growth rate | Absolute pressure | Gas flow rates | Growth setup | Feed |
|---|---|---|---|---|---|
| M1 | 50 mm/h | 7.5 bar | Ar: 3 L/min<br>$O_2$: 1 L/min | V.1 | arc-fusion type |
| M2 | 40 mm/h | 10.75 bar | $O_2$: 1.1 L/min | V.2 | arc-fusion type |
| M3 | 40 mm/h (main part)<br>20 mm/h (last part) | 7.5 bar | Ar: 0.75 L/min<br>$O_2$: 0.25 L/min | V.2 | arc-fusion type |
| M4 | 35-48 mm/h | 7.5 bar | Ar: 1.3 L/min<br>$O_2$: 0.33 L/min | V.2 | ceramic made from 5N powders (Roth Rotimetic) |
| M5* | 40-50 mm/h | 7.5 bar | Ar: 1.3 L/min<br>$O_2$: 0.33 L/min | V.2 | machined Kojundo ceramic |
| M6 | 40 mm/h | 7.5 bar | Ar: 1.3 L/min<br>$O_2$: 0.33 L/min | V.2 | machined Kojundo ceramic |

* For this exceptional case the growth direction changed from <100> to <110> directly after seeding

### 2.2 Thermodynamic calculations

Thermodynamic equilibria were calculated with the FactSage 8.1 software and database package (GTT Technologies, Herzogenrath, Germany 2021) for pure MgO and for MgO with oxidic impurities based on Al, Ca, Cr, Fe, Mn, Ti, V and Zn, for the atmospheres (composition and total pressure p) that were used in the crystal growth experiments.

### 2.3 Chemical analyses and microscopy

The chemical compositions of single crystalline samples were measured by (1) micro X-ray fluorescence spectroscopy (µ-XRF) and (2) inductively coupled plasma optical emission spectrometry (ICP-OES).

µ-XRF was used to characterize feeds, seeds and grown crystals to identify the most critical impurities. The measurements were performed under low vacuum conditions (below 1 mbar) using a Bruker M4 TORNADO spectrometer. For the point measurements, the rhodium X-ray source was operated at a voltage of 50 kV and the tube current was set to 600 µA to achieve maximum sensitivity. The samples were oriented in such a way that the Bragg peaks did not disturb the relevant energy regions of the spectra. Polycapillary X-ray optics were used to focus the white radiation at the surface of the sample, resulting in a spatial resolution of about 20 µm. The measurement time per point was set to 100 s. Furthermore, µ-XRF elemental mappings were carried out to investigate the chemical homogeneity of chemo-mechanically polished (CMP) sections of selected crystals. Each sample area was analyzed almost in its entirety with a measurement grid distance of about 15 µm using the "Area"-mode of the spectrometer. The surface areas were scanned "on the fly" by moving the sample stage continuously. The measurement time per point was set to 10 ms and the spots were measured 14 times to increase the counting statistics, i.e. fourteen passes of the scans were performed. To perform mappings with highest sensitivity, the Rh X-ray source was operated under the same conditions as for the point measurements. An additional mapping was performed for the arc-fusion grown feedstock material with a measurement grid distance of about 110 µm, a measurement time per point of 10 ms and with four passes of the scans. For all measurements the quantification of the detected elements was conducted using the fundamental parameter (FP) approach based on Sherman's equation (Sherman, 1955).

Standard-based inductively coupled plasma optical emission spectrometry (ICP-OES) was used as complementary technique to quantify the impurity levels of critical elements accurately. Additionally, it was necessary to quantify aluminum, which is not accessible by the µ-XRF spectrometer at relatively low quantities. The impurity levels were measured using a Thermo Scientific iCAP 7400 Duo Series ICP-OES device. The samples in the form of small crystal pieces were etched with concentrated $HNO_3$ for 10 minutes to remove any potential residues from the surfaces prior to microwave digestion in $HNO_3$ for 20 min at 220°C. The spectrometer was calibrated with matrix-adapted synthetic solution standards and each sample reading had three replicates.

Chemo-mechanical polishing was carried out by CrysTec (Berlin, Germany) to prepare high-quality substrates of various sizes and to identify defects by using Nomarski differential interference contrast (DIC) microscopy.

### 2.4 Numerical calculations

For the calculation of temperature and stress fields during the growth, a numerical model in Elmer FEM software (P. Raback) based on the Finite Element Method was developed. The geometry in the simulation principally consists of feed, molten zone and crystal as shown in Fig. 1. Very simplified shapes of feed and crystal holders were also added to account for conductive heat losses. The thermal model includes:

- Radiation losses to ambient: $Q_{rad} = \varepsilon \sigma_{sb}(T^4 - T_a^4)$, where $\varepsilon$ is surface emissivity (a value of 0.9 was assumed), $\sigma_{sb}$ is the Stefan-Boltzmann constant and $T_a$ is the ambient temperature (assumed 300 K).
- Irradiation approximated with a Gaussian profile (cf. (Souptel *et al.*, 2007)) on the material surface depending on the vertical coordinate (z): $Q_{in}(z) = Q_0 \, exp\left[-\frac{(z-z_0)^2}{2w^2}\right]$, where the amplitude $Q_0$ was adjusted to fit the melting point at the melt-crystal interface, while the width w = 4 mm and midpoint $z_0$ = 1.35–1.8 mm were adjusted to fit the molten zone height.
- Convective cooling by gas flow: $Q_{conv} = \alpha(T - T_a)$, where a heat transfer coefficient α = 8 W/m²K was estimated considering a forced flow over a cylindrical body (*Vdi-wärmeatlas*, 2006) inside the quartz tube. The gas mixture and flowrate from the experiment was applied.

The solid-liquid interfaces were fitted to the melting point isotherm (2825 °C) using a moving mesh approach while the melt free surface in the model was assumed flat and fixed. For the thermal conductivity of crystal, melt and feed, we applied $\lambda[W/mK](T[K]) = 63728 \cdot T^{-1.242}$ (Hofmeister, 2014). Since the crystal and melt can be partly transparent to heat radiation, the Rosseland model was also applied, which has been used in the literature for oxide crystals, e.g. in Stelian et al. (Stelian *et al.*, 2019). In this model, an effective thermal conductivity is defined: $\lambda_{eff}[W/mK](T[K]) = \lambda_{solid} + \frac{16n^2\sigma_{sb}T^3}{3a}$, with the refractive index *n* = 1.8 (Djaili *et al.*, 2020) and the absorption coefficient *a* = 25000 1/m (Chernyshev *et al.*, 1993). These values correspond to high temperatures close to the melting point and short wavelength of about 1 μm (radiation maximum according to Planck's spectrum at 2825 °C is at 0.9 μm). However, it should be noted that the absorption coefficient strongly depends on the impurities in the crystals and may be significantly smaller for highly pure crystals. Such a case was also considered in simulations.

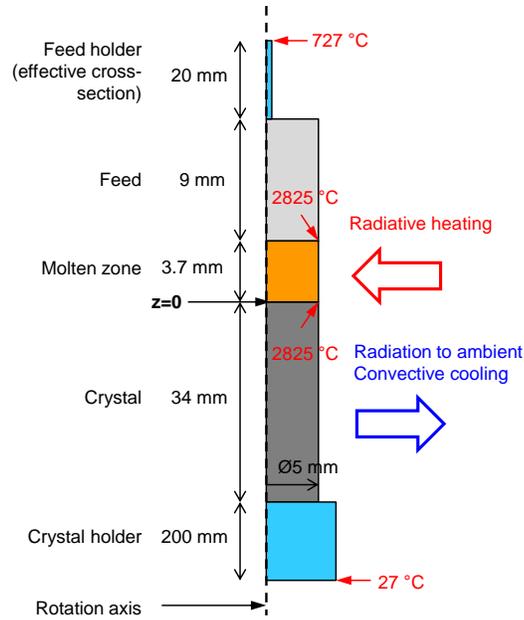

**Figure 1** Geometry and boundary conditions in the thermal simulation.

### 3. Results and discussion

#### 3.1 Purity considerations of feed and seed rods

Single-crystalline and polycrystalline feed rods with different purities were used for the single crystal growth experiments. Typical feeds and a ⟨100⟩-oriented seed are shown in Fig. 2. The arc-fusion grown feedstock shows an average purity of about 99.94 %, determined by large scale µ-XRF mappings and by ICP-OES investigations of a randomly selected cleaved sample. The chemical investigations of the ceramic and the powders revealed excellent purity, comparable to the nominal purity levels given by the suppliers (see Table 2).

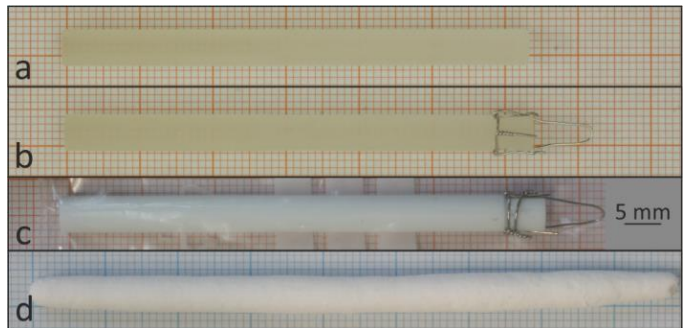

**Figure 2** Typical (a) seed and (b-d) feed rods used for single crystal growth of MgO. The shown feeds were (b) machined from an arc-fusion grown crystal, (c) prepared from a MgO ceramic with 4N5 purity and (d) made by the classical ceramic technique using 5N purity MgO powder.

**Table 2** Results of the chemical investigations of the starting materials performed by ICP-OES and µ-XRF in ppm by weight. The results of the latter technique are shown in brackets. Some of the ICP-OES values are below the limit of quantification (LOQ).

| sample | Al (ppm) | Ca (ppm) | Ti (ppm) | V (ppm) | Cr (ppm) | Mn (ppm) | Fe (ppm) | Zn (ppm) |
|---|---|---|---|---|---|---|---|---|
| Arc-fusion grown material for initial feeds and seeds | 81±0.8 | (313±66) | <4.8(14±3) | 36±0.43 (25±1) | 28±0.67 (72±18) | 13±0.16 (4±2) | 76±0.88 (72±2) | <4.2 (0-15) |
| *MgO ceramic target sample as delivered (manufactured by Kojundo, theoretical purity given by the manufacturer: 99.995 %)* | <10.8 | (29±1) | <4.8 | <9.3 | <12.3 | 8±0.07 | 8±0.35 | 11±0.1 |
| 5N (99.999%) MgO powder as delivered by Roth (ROTIMETIC) | <3.1-16 | <0.07-0.8 | <0.8 | <2 | <1 | <0.3 | <1.9 | <0.8 |

### 3.2 Thermodynamic calculations

Figs. 3a-d show the activity of gaseous species under different experimental conditions. It should be noted that for an ideal gas (which is a reasonable approximation under the given experimental conditions) the activity equals the vapor pressure in bar. Hence, an activity of 0.05 corresponds to a vapor pressure of 50 mbar, which is already so high that severe evaporation occurs.

Fig. 3a shows isothermal equilibria of pure MgO slightly above the melting point (2830°C) in atmospheres ranging from pure Ar (left) to pure $O_2$ (right). MgO can evaporate as molecule (MgO) as well as elemental Mg after dissociation. The first evaporation can be described by the physical process $MgO_{(sol/liq)} \longrightarrow MgO_{(gas)}$, and the equilibrium is only influenced by the temperature T. Because here T is constant, the fugacity does not depend on the total pressure or on the gas composition. However, this is not the case for evaporation as atoms via $MgO_{(sol/liq)} \longrightarrow Mg_{(gas)} + 0.5\ O_{2(gas)}$. This chemical equilibrium shifts to the product side for low oxygen concentrations. Hence, the evaporation as atoms prevails in atmospheres containing less than ca. 10% $O_2$. This effect becomes stronger for lower p, because then even less oxygen is present.

Fig. 3b and 3d compare on a temperature scale the fugacities of MgO and Mg (black dashed or dotted lines, respectively) with the fugacities of species which prevail as oxides of the impurity elements mentioned above. The metals themselves partially prevail (Zn, Mn), and sometimes oxides are found with different valences. The calculations were performed for constant p = 7.5 bar and Ar atmosphere containing 19.88 or 25% $O_2$. It turns out that this concentration difference has no strong impact, because the concentrations of gases enter only logarithmically into the equilibria. Zn (metal) and the oxides $VO_2$, $CrO_2$, and FeO reach high fugacities already for temperatures that are significantly lower than for MgO and Mg. Consequently, preferred evaporation of these impurities must be expected. The difference is much smaller for Mn, but this metal is still expected to evaporate more strongly than the host MgO. Because Mn also evaporates as metal, the evaporation temperature shifts to slightly lower T if less oxygen is present (the fugacity of MgO is not influenced, only that of Mg). Al, Ca and Ti do not form species with high fugacity, which impedes preferred evaporation from a MgO melt.

Fig. 3c shows a similar calculation for pure $O_2$ with higher total pressure p = 10.75 bar. The order of fugacities is similar, but the formation of $Mn_{(gas)}$, and also of $Mg_{(gas)}$ is so strongly suppressed that Mn evaporates less than MgO - with consequently negative impact on purification.

Generally, it should be noted that even for similar fugacities of MgO/Mg and an impurity, a purification effect can be expected because under a similar tendency to evaporate, depletion of the minority component can be expected.

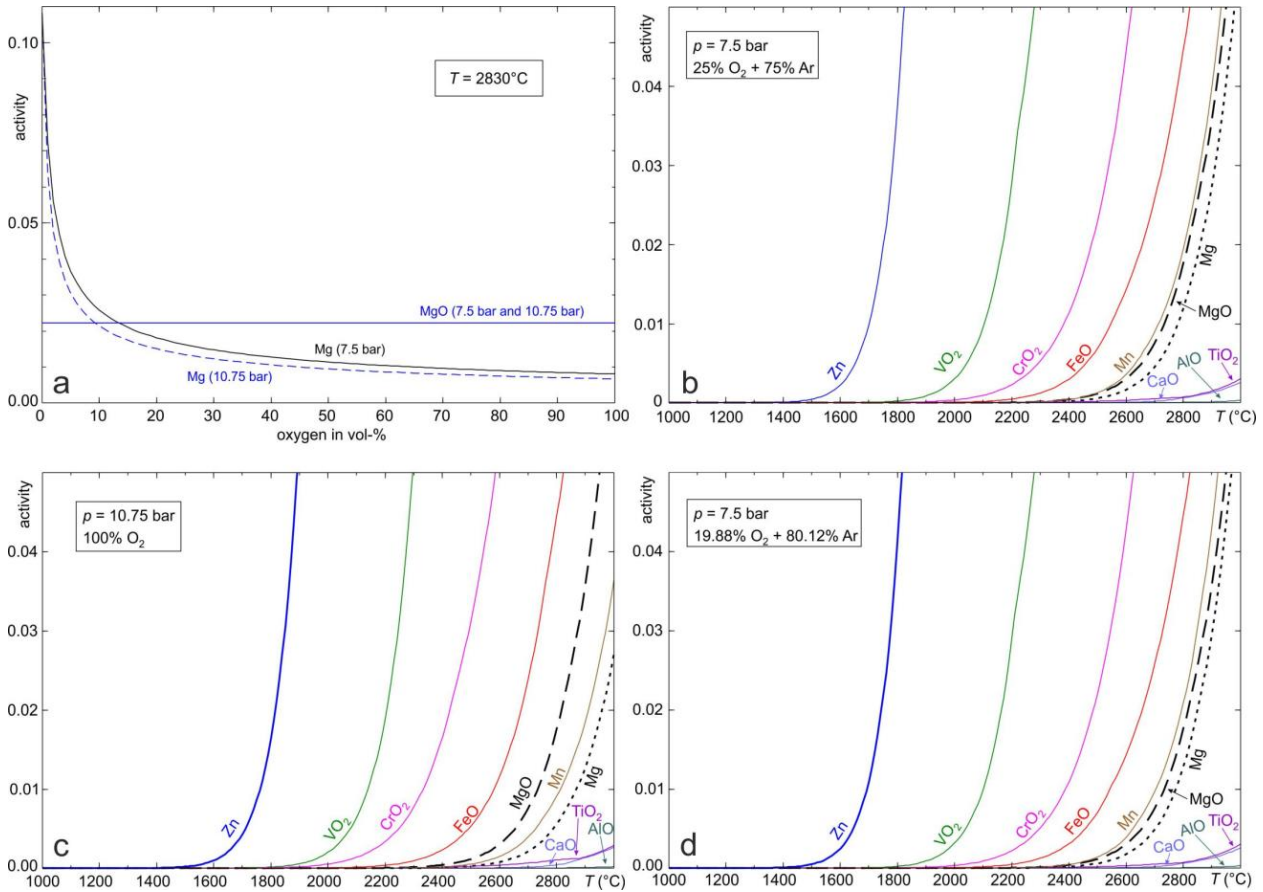

**Figure 3** (a) Fugacities of $Mg_{(gas)}$ and $MgO_{(gas)}$ at 2830 °C (slightly above the melting point) in atmospheres ranging from pure Ar (left) to pure $O_2$ (right). (b-d) Fugacities of MgO, Mg (black dashed or dotted lines, respectively) and impurity species on a temperature scale for the different growth atmospheres.

### 3.3 Crystal growth of MgO crystals

MgO single crystals with diameters between 3.5 and 5 mm and lengths between 10 and 40 mm were grown by the optical floating zone technique in a flow of Ar/$O_2$ gas mixtures or in pure $O_2$ at increased total pressures between 7.5 to 10.75 bar (see Table 1 for growth conditions).

The crystal lengths achieved in the first experiments with gas flow directed from the top to the bottom was strongly limited due to the condensation of MgO on the inner glass tube, reducing the available radiation to keep the zone in the molten state. Furthermore, vapor growth of MgO occurred at the surface of the melt-grown crystal, as can be seen in Figs. 4a-c. In some growth experiments (results are not shown), the seeding process was disturbed by the condensed material. This issue was eliminated by a directed gas flow from the bottom to the top of the chamber. Therefore, condensation of gaseous species at the surface area of the incoming radiation and at the grown crystal was drastically reduced. By the application of feed rods with lower purity, especially arc-fusion grown material, the growth was very stable, and no cleaving of the crystals occurred. The resulting crystals (M2 and M3) grown at different oxygen partial pressures are shown in Fig. 5.

Feed rods with much higher purity led to a drastic increase in optical transmission, resulting in less stable growth due to partial freezing of the molten zone at maximum lamp power and fully opened shutters. This was mainly observed for crystal growth experiments using feed rods prepared from ceramics with higher purity and density (Kojundo). Rods with lower density slightly improved the outcome, but crystal growth was still working at the limits of the growth station. It should be noted that the crystals withstood the harsh growth conditions and the increased thermal stress without cleaving. On one occasion a crystal had to be partially remelted; even the repositioning with an extreme rate of 1000 mm/h only led to near-surface cracks (see section below) in minor regions, where exceptionally high thermal stress occurred.

The observed growth instabilities can be explained by two major factors: (1) the increased purity of the crystals reduced the optical absorption of MgO in the wavelength range of the radiation source and (2) the purity also increased the thermal conductivity due to the reduction of scattering centers at residual impurities (Morton & Lewis, 1971). Similar challenges are typical for oxides with low absorption characteristics, such as corundum (α-$Al_2O_3$) at increased purity (Field & Wagner, 1968, Guguschev *et al.*, 2010).

Some of the high-purity crystals (Fig. 6) show an increased cloudiness. This peculiarity depends on the amount, size and distribution of voids as explained in detail in the following section.

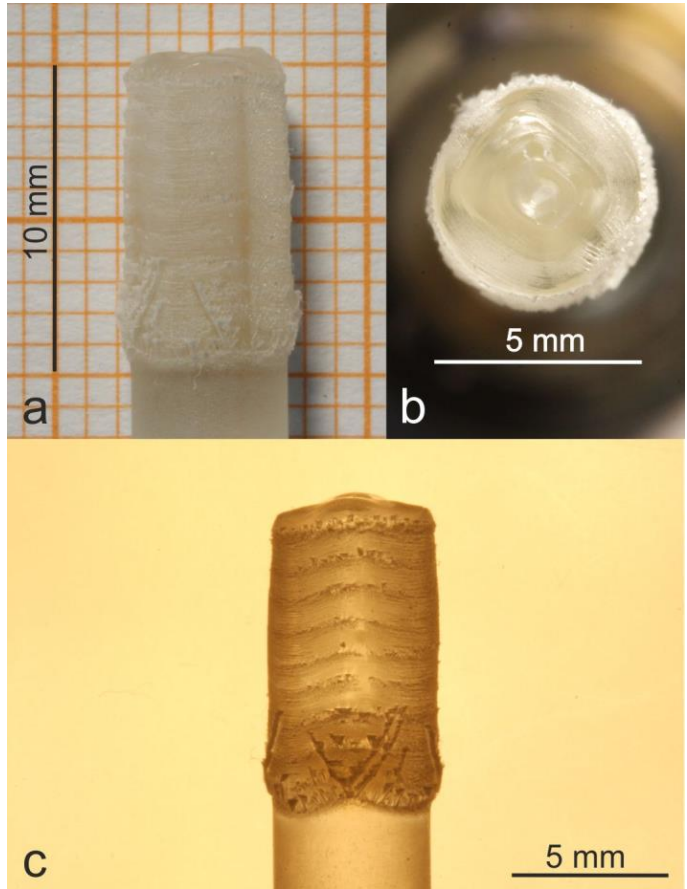

**Figure 4** First MgO crystal M1 (length: 10 mm, diameter: 5-6 mm) grown using the optical floating zone technique ((a) optical photograph on graph paper, (b) top view on the last solidified part, (c) optical photograph under transmitted light). The surface of the melt-grown crystal is partially covered by MgO grown from the vapor phase.

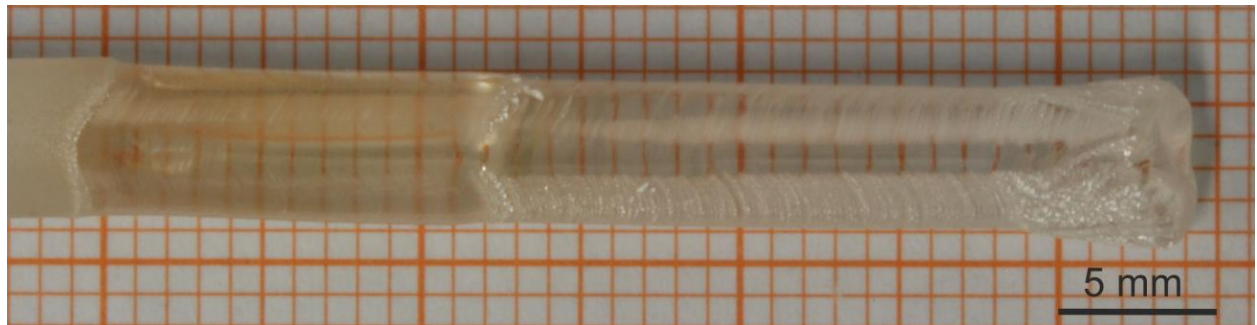

**Figure 5** MgO crystal M2/3 grown sequentially with a total length of about 35 mm (about 13 mm for M2 and 22 mm for M3). M2 shows a slightly yellowish tint and the second half of the boule is "water clear". The grid size is one mm by one mm.

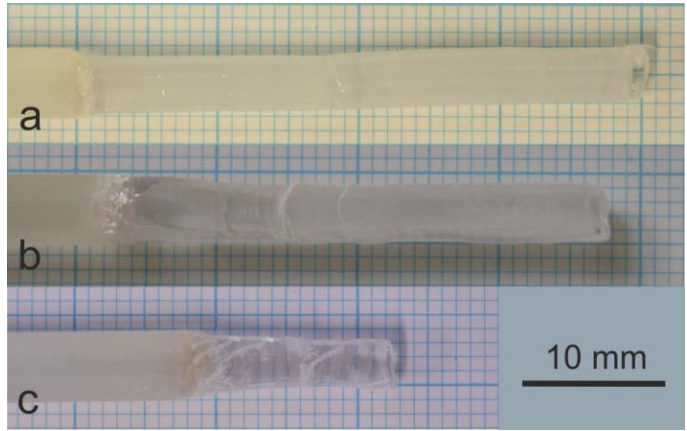

**Figure 6** High purity MgO crystals (a) M4, (b) M5 and (c) M6 with diameters between 3.5 and 4.5 mm and lengths between 14 and 40 mm.

### 3.4 Simulation results

Numerical simulation results of the MgO crystal growth process in Fig. 7 with a reference absorption coefficient $a = 25000$ 1/m indicate that the absorbed irradiation of 861 W is mostly balanced by radiation from the surfaces of the melt, crystal and feed. Heat convection in the gas as well as heat losses in the feed and crystal holders clearly play only a minor role. The solid/liquid phase boundaries demonstrate that the molten zone height decreases in the center, which is confirmed experimentally in Section 3.5. However, this effect may depend on the convection in the melt and the exact meniscus shape, which were not considered in the simulation. If the absorption coefficient is decreased by a factor of 10 down to $a = 2500$ 1/m (and irradiation profile is adjusted with $z_0 = 1.35$ mm), the absorbed power increases by 2%. A high purity feed material (with less infrared absorption expected) increased the required shutter opening (nonlinear relation to radiation power!) by about 15% in the experiment, which shows the same trend.

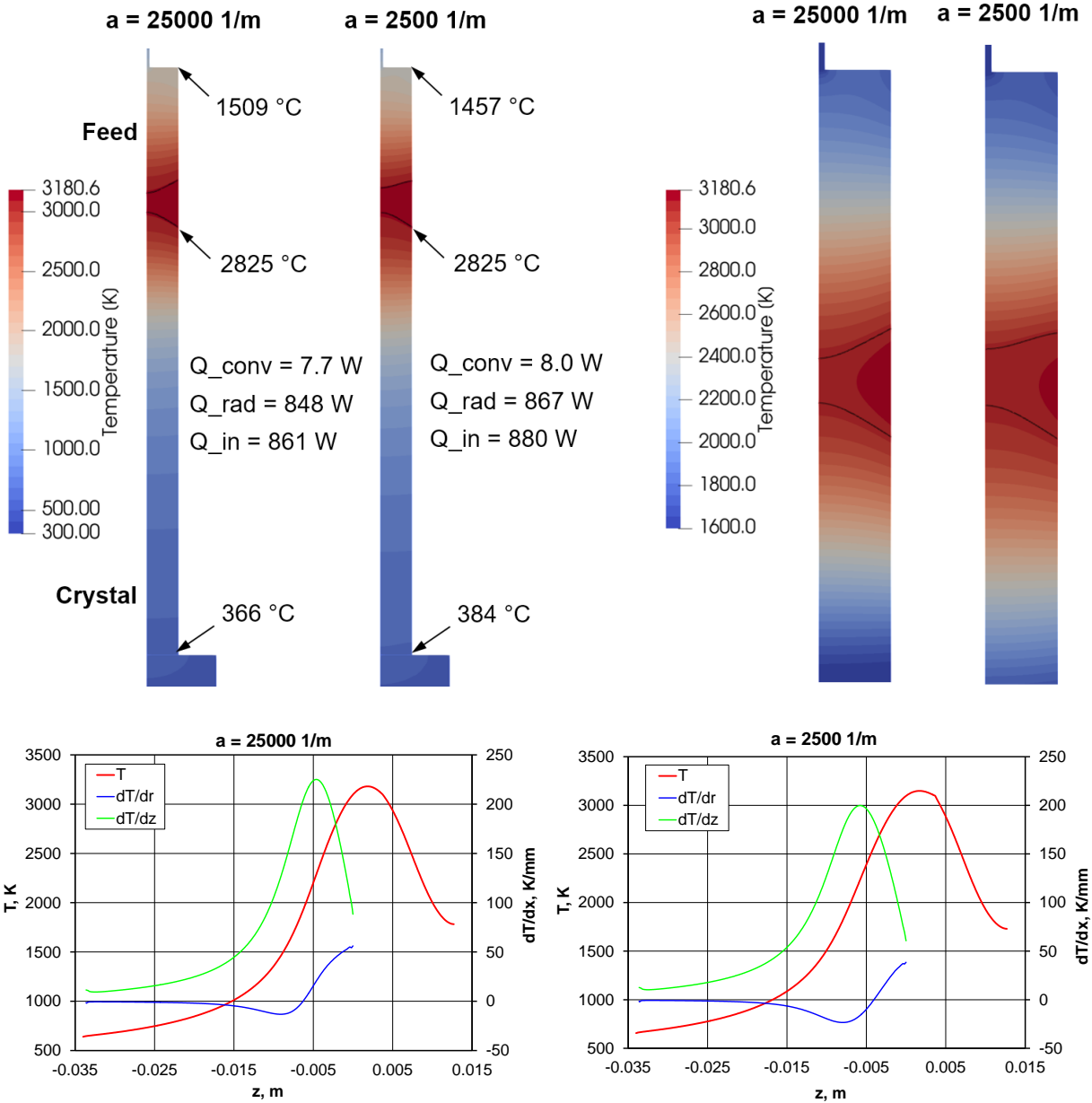

**Figure 7** Simulation results: temperature field (top, zoom-in on the right) and distributions on the material surface (bottom).

The temperature distribution on the crystal surface in Fig. 7 has a symmetric shape with the maximum temperature in the melt, a decrease down to 400 °C in the crystal and 1500 °C in the feed. It was observed in an experiment that the feed holder made of platinum wire was partially molten during the experiment. The melting point of platinum of 1768 °C is reached at about 3 mm from the end of the 9 mm long feed used in the simulation, which qualitatively confirms this observation.

Thermal gradients on the crystal surface reach 225 K/mm in the axial direction and over 50 K/mm in the radial direction (see Fig. 7). A change of sign of the radial gradient can be observed between 4 to 7 mm below the triple point when the absorbed heat becomes smaller than the radiation losses. The gradients are generally lower with a smaller absorption coefficient (larger effective thermal conductivity).

With the known temperature distribution in the crystal, the thermal stress was calculated assuming an isotropic material model. The elastic moduli were calculated as isotropic Voigt-Reuss-Hill averages using single-crystal data from (Isaak *et al.*, 1989) and a linear approximation in the temperature range 800-1800 K: $E[Pa] \approx 2.85 \cdot 10^{11} - 5.564 \cdot 10^7 (T[K] - 800)$ and $G[Pa] \approx 1.19 \cdot 10^{11} - 2.475 \cdot 10^7 (T[K] - 800)$. An approximation for the linear thermal expansion coefficient was derived from data in (Suzuki, 1975) as $\alpha[1/K] \approx 10^{-6}(12.4 + 0.0025 T[K])$. The calculated stress field in Fig. 8 shows a maximum von Mises stress of 221 MPa. The maximum stress is located in the region with the highest radial temperature gradients in the top part of the crystal. A lower absorption coefficient (higher crystal purity) decreases the thermal stress at stable growth conditions (growth stability is discussed in Section 3.3), because the thermal gradients become lower. The critical resolved shear stress (CRSS) for MgO at temperatures over 500 K is below 20 MPa according to (Amodeo *et al.*, 2018), so that the thermal stress may be relaxed by dislocation multiplication during the process.

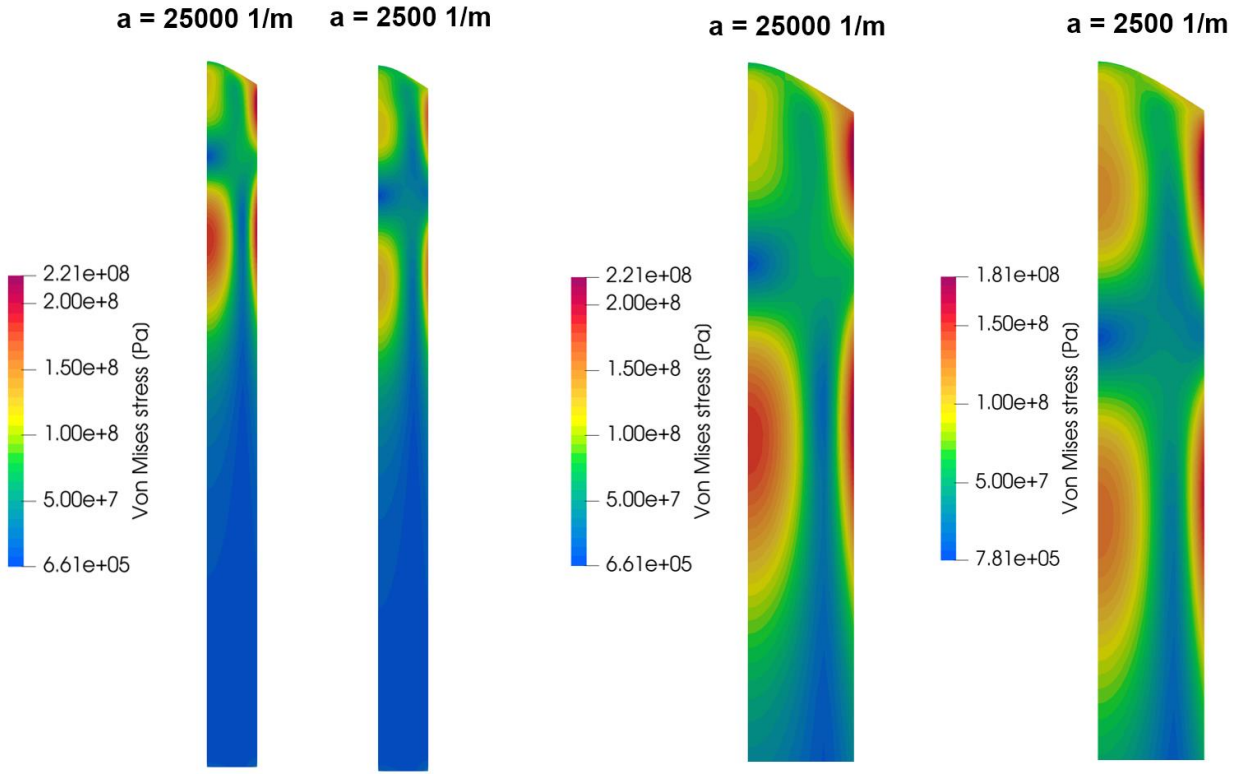

**Figure 8** Simulation results: thermal stress field (von Mises) in the crystal. Zoom-in on the top crystal part is shown on the right.

### 3.5 Substrate preparation and DIC microscopy

DIC micrographs of the substrate crystals M2-3 (Fig. 9) revealed that the floating zone grown MgO crystals contain voids. Most of the clearly observable voids appear as cuboid-shaped negative crystals, i.e. as miniaturized cavities surrounded by crystallographic {100} planes. It is noticeable that the appearance and distribution of these voids depend on the applied oxygen partial pressure and the used feed rods. The crystal M3 was grown at an oxygen partial pressure $pO_2$ of about 1.875 bar and the formed voids only appeared in the central part of the crystal. This is in clear contrast to the section grown at 10.75 bar $pO_2$, were voids were additionally detected close to the rim of the crystal. DIC micrographs of the high-purity crystals grown from ceramic rods at a $pO_2$ of about 1.5 bar (M4-6) are shown in Fig. 10.

A very detailed work about the formation of voids in arc-fusion grown MgO crystals was performed by Briggs in 1975 (Briggs, 1975a, b). The author pointed out that for MgO crystals, an equilibrium concentration of $OH^-$ ions remains in solution in the molten state, and is incorporated in crystals when the melt solidifies. This takes place even though most of the hydroxyl, usually present as $H_2O$, $Mg(OH)_2$ or adsorbed OH, is driven off, decomposed or reacted. The presence of magnesium vacancies and $OH^-$ ions in the crystal, and the reducing atmosphere above it, allows the formation of cavities, and liberation of hydrogen in them. Or more precisely: according to his model, oxygen vacancies and electrons combine with hydroxyl/magnesium vacancy complexes to form MgO divacancies and hydrogen. This type of defect diffuses and condenses with similar entities to form a void nucleus containing hydrogen gas. When all available vacancies and cation impurities have formed complexes with OH$^-$ ions, the remaining hydroxyl impurity is able to enter the MgO lattice as free, substitutional OH$^-$ ion.

According to this model, hydrogen/divacancy agglomerates which form close to the surface of the crystal are able to diffuse out, and this accounts for the absence of cavities in the outer regions of the crystals. This could be enhanced by vacancy diffusion along dislocation lines (otherwise known as fast pipe diffusion), since activation energies are lower compared to defect-free regions (Zhang *et al.*, 2010). Such effects could explain why we also do not see voids close to the periphery of the crystals grown at lower oxygen partial pressure (see Figs. 9-10).

As the oxygen partial pressure increases, the effective evaporation of MgO-related gaseous species (MgO$_{(gas)}$ and Mg$_{(gas)}$, see above in Section 3.2) is reduced and the oxygen vacancy formation energy also increases, which will lead to fewer oxygen vacancies (Yamamoto & Mizoguchi, 2013). Therefore, the probability that oxygen vacancies and electrons combine with hydroxyl/magnesium vacancy complexes to form MgO divacancies and hydrogen is reduced. On the other hand, to some extent the increased total pressure could hamper the out-diffusion of the formed hydrogen and/or other gases. These effects could explain why we observe fewer (but larger) voids in the interior and also closer to the periphery of floating zone grown crystals at high pO$_2$. Additional post growth annealing effects influencing the resulting void size cannot be excluded for crystal M2 since it served as the seed for crystal M3.

Furthermore, under our growth conditions, the formation of voids does not exclusively need to be as complex as described above: on the one hand, the formation of magnesium and oxygen vacancies might be sufficient to form divacancies, which serve as nuclei for voids (in this particular scenario: empty space) and increase their size by continuous attachment of further vacancies. This would be very similar to the postulated vacancy clustering processes for SrTiO$_3$ (Kok *et al.*, 2016).

On the other hand, as mentioned above, rapid cooling of the crystal introduces dislocations by thermal stress, which migrate by gliding and climbing to form subgrain boundaries and dislocation tangles. According to Briggs (Briggs, 1975a), these defect types serve as nucleating centers of the voids and at lower impurity levels the dislocations are more mobile due to less pinning at impurities. From the work on polycrystalline MgO it is also known that even a small quantity of pores of very small diameter can enhance the sliding by providing the source for intergranular sliding (Day & Stokes, 1966), which would result in a self-enhancing process. The porosity of the feeds can also cause an increased amount of dissolved and captured residual gases in the melt. So far, preliminary FTIR results of double-sided chemo-mechanically polished samples (with relatively thin sample thicknesses of 0.5 mm) measured at a pressure of about 7 mbar revealed no indication of OH$^-$ bands in clear and milky parts. Therefore, the mechanism in the formation of the observed voids in optical floating zone grown crystals could not be conclusively clarified.

For some of the crystals, glide bands were visibly pronounced when examined with DIC microscopy (see crystals M4 and M6 in Fig. 10). The formation of such glide bands is typical for plastically deformed MgO single crystals (Weaver & Paterson, 1969, Auten *et al.*, 1976). In the most extreme cases, small cracks filled with dendrites were formed by fast solidified melt in minor regions at the periphery of some high-purity crystals (see crystal M4 in Fig. 10 in the center of the top row). Their formation can be primarily explained by the highest thermal stress in these crystals being induced by abrupt changes in growth behavior (see above in Section 3.3). The nebulous appearance of certain sections in crystal M6 are characteristic for the parts where growth instabilities occurred. These DIC photographs (and some images for M4) validate the formation of a convex growth interface shape (see Fig. 10 in the center and the left of the top and bottom rows).

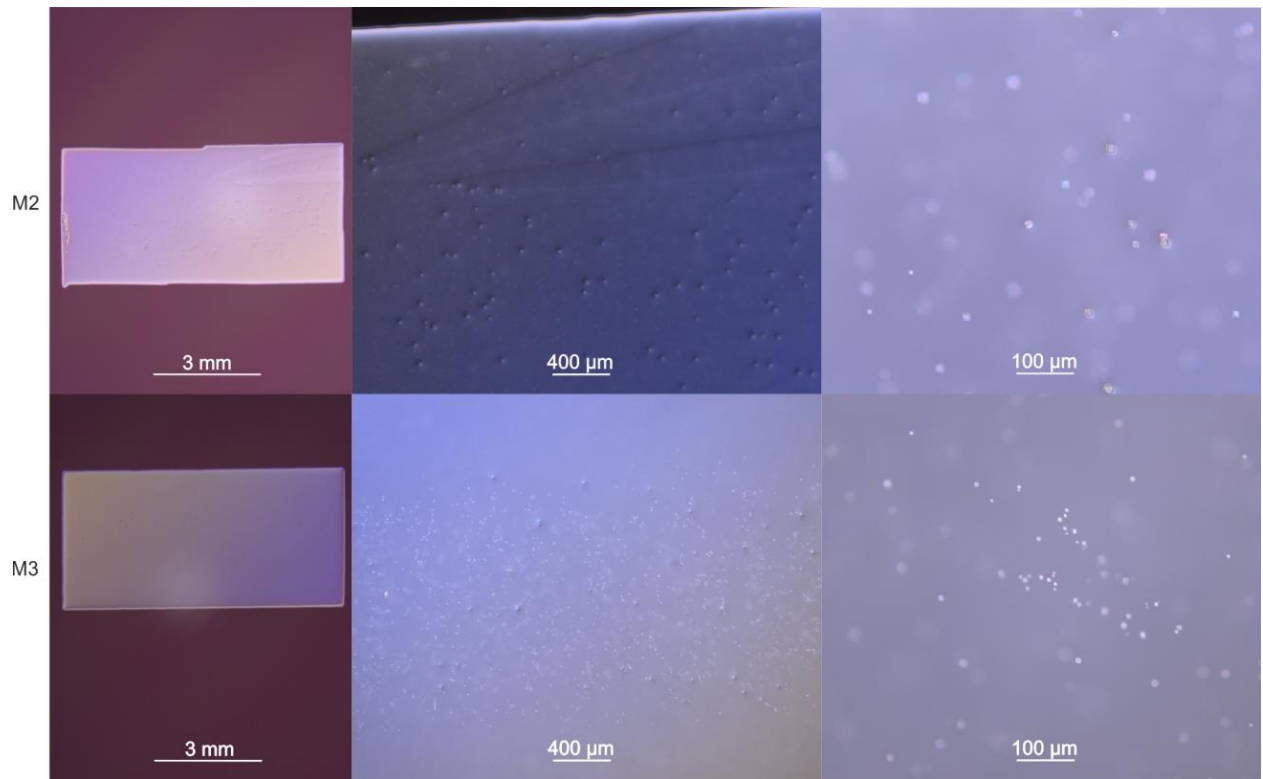

**Figure 9** Selection of DIC micrographs of polished sections of the crystals M2 and M3 at different magnification. The substrate of crystal M2 contains two visible subgrains, which grew out at a later growth stage.

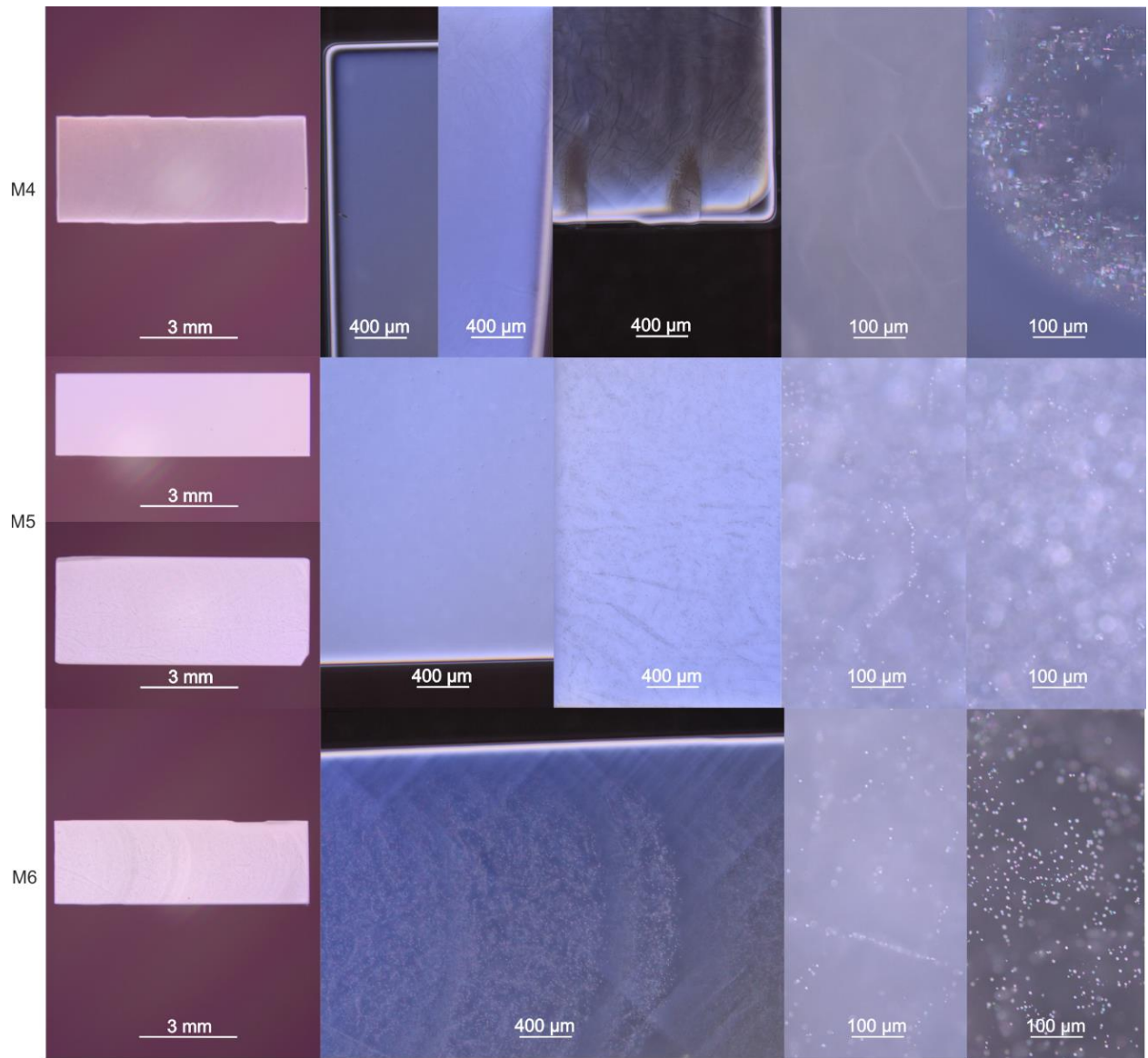

**Figure 10** Selection of DIC micrographs of polished sections of the crystals M4, M5 and M6 at different magnification. Voids and traces of glide bands are the dominating defect types in the grown high-purity crystals.

### 3.6 Chemical analyses of the grown crystals

The comparison of the impurity levels of the starting materials with the impurity concentrations of the grown crystals revealed that the applied technique is very effective in terms of purification. Due to the very high temperatures of the melt and the relatively large surface to volume ratio, the more volatile contaminants like Zn, V, Cr, Fe and Mn are effectively removed from the melt. This is clearly visible in Fig. 11 by comparing the impurity levels of the seed region with the selected region of crystal. This crystal was grown from feed material originating from arc-fusion grown crystals. Effective lowering of the Zn content can be seen by comparing the Zn value of the ceramic-based feed with the grown crystal (Tables 2, 5 and 6). Therefore, for these impurities, segregation plays a minor role in terms of purification. From the thermodynamic point of view, based on calculations (see above in Section 3.2) Ca, Al, and Ti are more critical due to having the lowest partial pressures of the evaporating species. The potential impact from Ti was largely negated due to negligible contamination levels. A drastic reduction of the Ca content was observed in our grown crystals, which indicates that under our conditions the effective evaporation could be not negligible and/or segregation is playing a role.

The most critical contaminant by far was Al, which mainly evaporates as AlO and exhibits the lowest vapor pressure. The results shown in Tables 2 and 4 indicate that the effective partition coefficient is smaller than 1, with a $k_{eff}$ value close to 0.56, but further studies are needed to verify this preliminary result. Nevertheless, according to the data obtained for the crystals M2 and M3 (Table 4), zone refining seems to be effective for this element, but at high initial values (as used for the arc-fusion feed material) it would require many passes to purify it to an acceptable concentration. This is why it is very important to take care that the initial feed rods do not contain aluminum in higher quantities. Therefore, the crystals grown from high-purity ceramics are superior with respect to purity compared to the single crystalline feed material (see Fig. 12 and Tables 5-6), since the initial concentration of contaminants is already very low and the purification effect does not need to be particularly effective.

In the demonstrated crucible-free technique, there are no electrodes or parts from the growth setup that could pollute the melt with contaminants as is the case in the arc-fusion technique. The slightly increased values of some of the impurities compared to the high-purity starting materials most likely originate from the seed crystals used for growth, since any potential residues at the surfaces of the samples (e.g. caused by cutting, cleaving or polishing procedures) were removed by $HNO_3$ etching prior to the ICP-OES measurements.

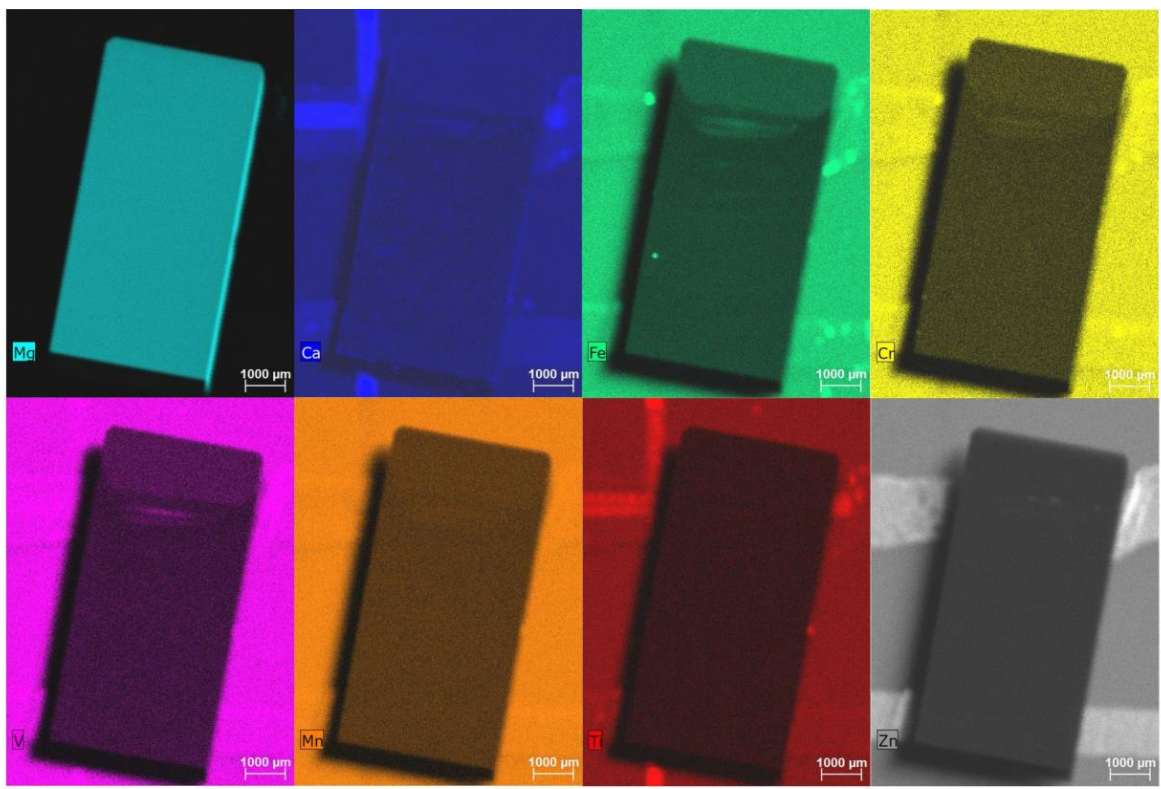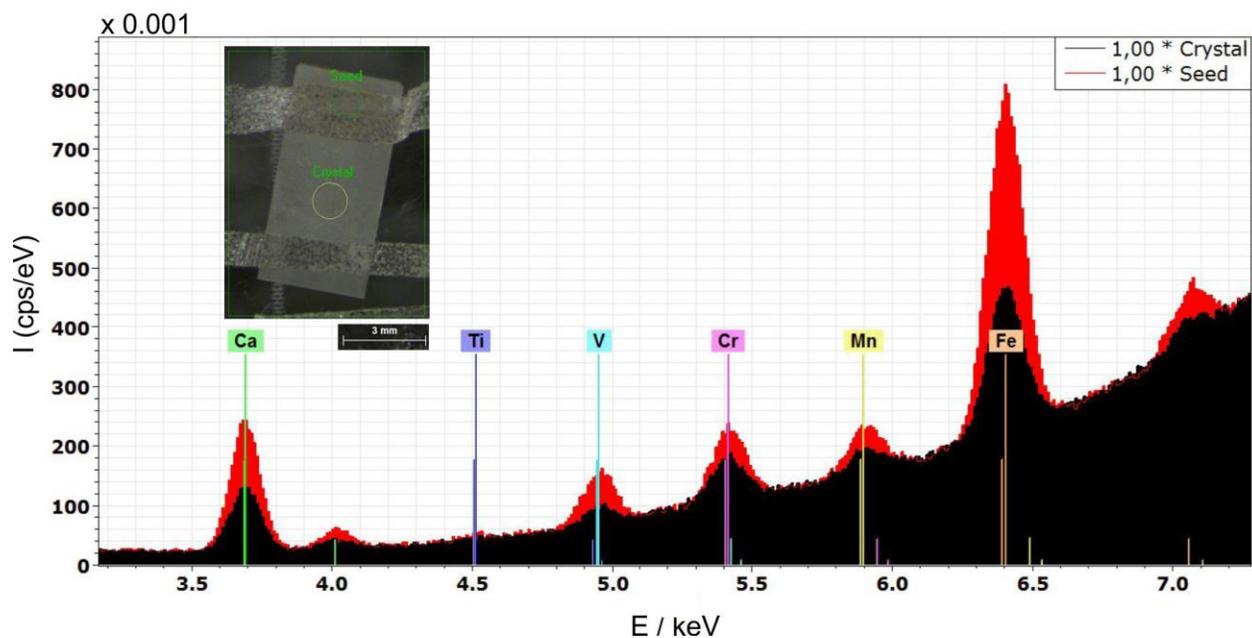

**Figure 11** Selected µ-XRF element mapping results for Mg, Ca, Fe, Cr, V, Mn, Ti and Zn for a polished substrate sample (top rows). It is evident that a fraction of the seed used is part of the polished substrate. The spectrum (bottom) measured at the seed region is compared with a spectrum of the grown crystal. The locations of the sum spectra of the different areas (crystal, seed) are indicated in the inset.

**Table 3** µ-XRF results of the investigated substrate (crystal M1 grown from an arc-fusion feed) in ppm by weight.

| Spectrum | Ca (ppm) | Ti (ppm) | V (ppm) | Cr (ppm) | Mn (ppm) | Fe (ppm) | Zn (ppm) |
|---|---|---|---|---|---|---|---|
| Seed | 186 | 1 | 25 | 22 | 10 | 75 | <LOD |
| Crystal | 90 | 0 | 6 | 13 | 3 | 30 | <LOD |

**Table 4** Results for the crystals M2 and M3 (grown from arc-fusion feeds) of the chemical investigations performed by ICP-OES and µ-XRF in ppm by weight. The results of the latter technique are shown in brackets.

| sample | Al ppm | Ca ppm | Cr ppm | Fe ppm | Mn ppm | Ti ppm | V ppm | Zn ppm |
|---|---|---|---|---|---|---|---|---|
| M2 (shortly behind the seed) | 45±0.9 | (198±12) | 15±0.07 (26±1) | 36±0.09 (94±4) | 6±0.053 (11±2) | <4.8 (3±1) | 13±0.42 (32±1) | <4.2 (<LOD) |
| M2 (close to the end of the crystal) | 68±0.53 | (95±3) | <12.3 (5±1) | 39±0.16 (31±0) | 4±0.05 (3±2) | <4.8 (0) | <9.3 (3±0) | <4.2 (<LOD) |
| M3 (XRF: 2-3 mm behind the seed ICP-OES: 4-5 mm behind the seed) | 97±0.9 | (116±10) | <12.3 (9±1) | 21±0.45 (25±2) | 3±0.154 (2±2) | <4.8 (2±2) | <9.3 (4±2) | <4.2 (n.d.)* |
| M3 (15-16 mm behind the seed) | - | (147±26) | (8 ± 0) | (24±2) | (2±1) | (1±1) | (5±1) | (<LOD) |
| M3 (close to the end of the crystal) | 300±0.8 | (138±5) | <12.3 (6±0) | 26±0.38 (23±1) | 3±0.036 (1±1) | 8±0.07 (1±1) | 11±0.37 (6±2) | <4.2 (n.d.)* |

* not detectable due to the appearance of an interfering Bragg peak

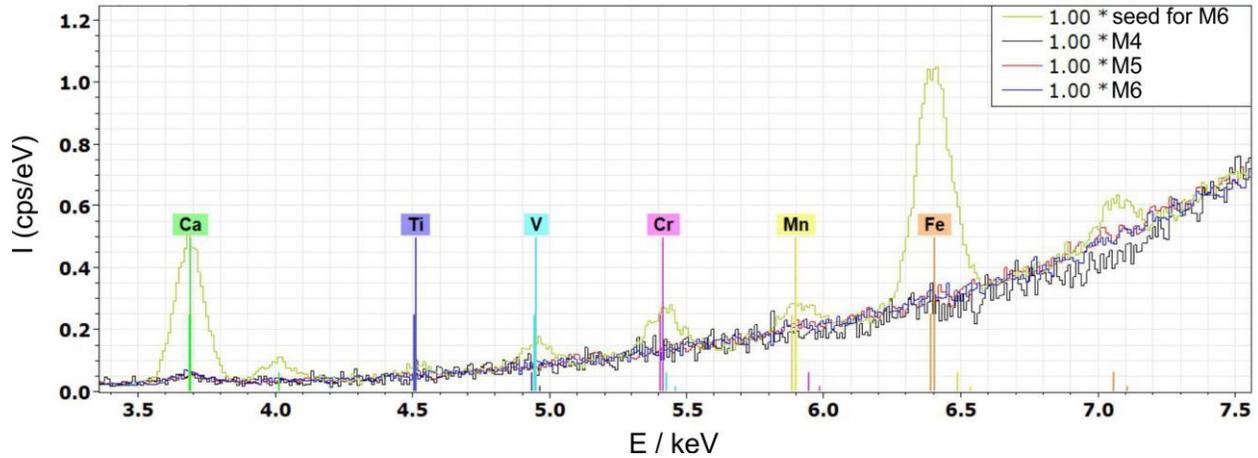

**Figure 12** µ-XRF spectra of a randomly selected MgO seed material in comparison with several crystals (M4, M5 and M6).

**Table 5** µ-XRF results of the investigated crystals M4, M5 and M6 for the first parts of the grown crystals shortly behind the seeds (in ppm by weight). The composition of a seed crystal (for M6) is shown for comparison. Most of the measured elements are below the detection limit (LOD).

| Spectrum | Ca (ppm) | Ti (ppm) | V (ppm) | Cr (ppm) | Mn (ppm) | Fe (ppm) | Zn (ppm) |
|---|---|---|---|---|---|---|---|
| Seed for M6 | 457 | 9 | 21 | 28 | 11 | 113 | <LOD |
| M4 | 11±2 | <LOD | <LOD | <LOD | <LOD | <LOD | <LOD |
| M5 | 19±1 | <LOD | <LOD | <LOD | <LOD | <LOD | <LOD |
| M6 | 25 | <LOD | <LOD | <LOD | <LOD | <LOD | <LOD |

**Table 6** Quantitative ICP-OES results for the investigated samples of M4, M5 and M6 (in ppm by weight). Most of the measured elements are below the limit of quantification (LOQ).

| sample | Al (ppm) | Ca (ppm) | Ti (ppm) | V (ppm) | Cr (ppm) | Mn (ppm) | Fe (ppm) | Zn (ppm) |
|---|---|---|---|---|---|---|---|---|
| M4 (at about 1/3 of the crystal length) | 10±0.03 | 6 | <1 | <2 | <1 | <0.3 | 8±0.40 | <0.8 |
| M5 (at about 1/3 of the crystal length) | <3.1 | 9±0.17 | <1 | <2 | <1 | 2±0.05 | 10±0.23 | <0.8 |
| M6 (close to the end of the crystal) | <3.1 | 5±0.02 | <1 | <2 | <1 | 0.7±0.07 | <2 | <0.8 |

## 4. Conclusion and outlook

In this study it was shown that the achieved MgO single crystals with lengths up to 40 mm and diameters between 3.5 and 5 mm are suitable for the preparation of high-purity single crystalline substrates. Due to the application of the optical floating zone technique, the state-of-the-art threshold of growing 4N purity crystals was exceeded by one order of magnitude, i.e. crystals with purities up to 5N were grown. Despite challenging material properties such as the high melting point of 2825 °C, very high evaporation rate and perfect {100} cleavage planes, crack-free crystals were grown at high growth rates in excess of 40 mm/h and at high thermal gradients.

The purest MgO single crystals were achieved by using high-purity ceramic feeds rods, a high flow of $Ar/O_2$ process gas mixtures (Ar: 1.3 L/min, $O_2$: 0.33 L/min), an increased absolute pressure of 7.5 bar and growth rates between 35 to 50 mm/h.

A simplified 2D thermal simulation of the presented growth process indicates high thermal gradients in the crystal of up to 225 K/mm in the axial direction. The resulting thermal stress reaches 220 MPa, and it decreases in high purity crystals due to higher effective thermal conductivity and hence lower thermal gradients. The thermal stress may be relaxed by plastic deformation due to the formation of dislocations, which migrate by glide and climb to form microstructural features such as glide bands, subgrain boundaries and dislocation tangles. Such microstructural features were also observed for arc-fusion grown MgO crystals and the appearance of subgrains is still common in commercial MgO substrates (Schroeder *et al.*, 2015).

The next steps for high-purity MgO crystals grown by the optical floating zone technique will be focused on stabilizing the growth process by using Xenon short-arc lamps with an increased output power, which compensates for the radiation loss as a result of the increased infrared transmission and thermal conductivity of the crystals. This would allow detailed studies of the segregation behaviour of Ca and Al at low concentrations and also a second pass of the melt zone towards the end of the feed rod to increase the purity of the crystal further. To investigate the origin of the observed voids and to reduce their density in high-purity crystals, further growth experiments and FTIR investigations with larger sample dimensions are necessary. Due to the relatively low density of MgO, X-ray micro-computed tomography (micro-CT) could be used to study the void distribution and density in detail for crystals grown under various conditions. In addition, detailed measurements of the temperature-dependent optical absorption coefficients in the near infrared spectral region would be useful for more accurate thermal simulations and process optimizations.

**Acknowledgements** The authors thank S. Kalusniak and S. Püschel for performing FTIR measurements and R. Blukis and C. Berryman for proofreading the manuscript. K. D. has received funding from the European Research Council (ERC) under the European Union's Horizon 2020 research and innovation programme (Grant Agreement No. 851768). A.P. gratefully acknowledges financial support from Dario Gil, Director - IBM Research for this study.

**Conflicts of interest** There are no conflicts of interest.

**Data availability** Our research data includes sensitive or confidential information. Therefore, some data may need to be requested and approved by IBM and IKZ.

**Author contributions**

C. Guguschev: methodology, formal analysis, investigation, writing original draft, visualization, funding acquisition, resources

M. Schulze: investigation, methodology, writing - review & editing

A. Dittmar: investigation, methodology, writing - review & editing

D. Klimm: investigation, methodology, writing original draft

K. Dadzis: software, investigation, methodology, writing original draft

T. Schroeder: writing - review & editing, project administration, resources

K. Peters: investigation, methodology, resources, writing - review & editing

A. Pushp: conceptualization, supervision, resources, project administration, funding acquisition, writing - review & editing